\documentclass[conference]{IEEEtran}
\IEEEoverridecommandlockouts

\usepackage{cite}
\usepackage{amsmath,amssymb,amsfonts}
\usepackage{graphicx}
\usepackage{algorithm}
\usepackage{algpseudocode}
\usepackage{textcomp}
\usepackage{listings}
\usepackage{xcolor}

\usepackage[
singlelinecheck=false 
]{caption}
\lstdefinelanguage{sparql}{
  morekeywords={
    ASK,SELECT,CONSTRUCT,DESCRIBE,WHERE,FROM,PREFIX,
    LIMIT,OFFSET,FILTER,OPTIONAL,GRAPH,INSERT,DELETE
  },
  sensitive=true,
  morecomment=[l]{\#},   
  morestring=[b]"
}

\def\BibTeX{{\rm B\kern-.05em{\sc i\kern-.025em b}\kern-.08em
    T\kern-.1667em\lower.7ex\hbox{E}\kern-.125emX}}
\begin{document}

\title{TRUCE: TRUsted Compliance Enforcement Service for Secure Health Data Exchange}

\author{Dae-young Kim and Karuna Pande Joshi, Senior Member IEEE
\IEEEcompsocitemizethanks{\IEEEcompsocthanksitem D. Kim and K.P. Joshi are with the Department of Information Systems, University of Maryland, Baltimore County, Baltimore MD 21227, USA
E-mail: \{leroy.kim, karuna.joshi\}@umbc.edu}
\thanks{UMBC Tech Report October 2023}}

\maketitle

\begin{abstract}
Organizations are increasingly sharing large volumes of sensitive Personally Identifiable Information (PII), like health records, with each other to better manage their services. Protecting PII data has become increasingly important in today’s digital age, and several regulations have been formulated to ensure the secure exchange and management of sensitive personal data. However, at times some of these regulations are at loggerheads with each other, like the Health Insurance Portability and Accountability Act (HIPAA)  and Cures Act; and this adds complexity to the already challenging task of Health Data compliance. As public concern regarding sensitive data breaches grows, finding solutions that streamline compliance processes and enhance individual privacy is crucial. We have developed a novel TRUsted Compliance Enforcement (TRUCE) framework for secure data exchange which aims to automate compliance procedures and enhance trusted data management within organizations. The TRUCE framework reasons over contexts of data exchange and assesses the trust score of users and the veracity of data based on corresponding regulations. This framework, developed using approaches from AI/Knowledge representation and Semantic Web technologies, includes a trust management method that incorporates static ground truth, represented by regulations such as HIPAA, and dynamic ground truth, defined by an organization’s policies. In this paper, we present our framework in detail along with the validation against the Health Insurance Portability and Accountability Act (HIPAA) Data Usage Agreement (DUA) on CDC Contact Tracing patient data, up to one million patient records. TRUCE service will streamline compliance efforts and ensure adherence to privacy regulations and can be used by organizations to manage compliance of large velocity data exchange in real time.
\end{abstract}


\section{Introduction}
Secure sharing of health datasets requires mitigating the risk of privacy and security breaches or getting contaminated and abnormal data. To minimize the risks, it must be possible to assess the trustfulness of data exchange parties and evaluate the veracity of the data. At the same time, the method must not interfere with eligible data exchange between harmless parties and fair data. This can be achieved by a system that reasons over various data access contexts and regulations. However, there is a lack of trust and veracity management framework. Researchers have explored the characteristics of trust and how to make it warranted with well-grounded evidence in the information technology area. What makes systems, users, or agents on behalf of users trustworthy? To which aspect of trust do we have to pay attention? If we can answer the first two questions, how do we derive components of trust from the user-system interaction? We strived to answer these questions, predominantly in mission-critical real-time data exchange \cite{kim2022mats}.

Trust matters more than ever in the current information systems because they coordinate dynamic and complex interactions of multiple users from numerous organizations in various environments. In most cases, the interactions involve large volume and high-velocity data exchange to achieve the system's goal. For example, health data exchange, such as contact tracing to reduce epidemiological risk \cite{kim2021trusted, kim2021semantically}, and disaster relief, such as search and rescue (SAR) \cite{kim2022mats}, are representative cases. In such systems, privacy or even lives are at stake when the systems fail to prevent unreliable data exchanges.

At the same time, the spread of fake news and misinformation prevails everywhere, and assessing the veracity of information is becoming increasingly difficult. Accordingly, there are deep learning studies on data veracity assessment to overcome this, which have achieved high accuracy. For example, Hardalov et al. distinguished fake news between 75\% and 99\% accuracy \cite{hardalov2016search}. Igawa et al. matched account patterns with humans and robots of Tweets to detect fraud with between 94\% to 100\% accuracy \cite{igawa2016account}. 

However, it would be challenging to integrate them into the real-time data exchange despite the performance of the algorithms. First, deep learning models need to understand the concepts of veracity and trust truly. Instead, they look for the combinations of parameters and weights of the convolutions that produce the best performances, mainly from the pre-labeled and pre-processed data input in a preferable format. As a result, it is difficult to know if the model creates inaccurate correlations between data features when it performs well.

The "Husky vs. Wolf" classifier experiment \cite{ribeiro2016should} gives us insights into the problem. Ribeiro et al. developed a wolf and husky classifier with 80 percent accuracy but focused on the undesirable correlation between features. They trained the wrong classifier on purpose to see if subjects could detect what was wrong inside it. The model focuses on the picture's background to classify images: a wolf if a picture has snow in the background; otherwise, it is a husky. It turned out that one-third of the subject trusted the model before they became aware of snow as a prominent feature.

The original purpose of the experiment was to emphasize the importance of the explainability of the model, but being able to explain after the incident is not acceptable in data exchange, especially for sensitive data. The same phenomenon will occur in data exchange if we train models to evaluate the trustworthiness of users and the veracity of data without providing concepts of trust. There must be a latent risk we cannot find before it occurs. Furthermore, the latent risk will produce severe privacy and security harm because data breach is irreversible.

Therefore, we propose a preemptive method to evaluate users' trust scores and the data's veracity. This paper expands our trust ontology \cite{kim2021trusted, kim2022mats} to include data veracity and solid validation with an US Centers for Disease Control and Prevention (CDC) Interim Guidance on Developing a COVID-19 Case Investigation \& Contact Tracing Plan \cite{cdcguide} and Health Insurance Portability and Accountability Act (HIPAA) Data Usage Agreement (DUA).

The overview of the paper is as follows. Section 2 discusses related work on trust components, especially on veracity. Section 3 illustrates our trust management framework. Sections 4 to 10 describe CDC Contact Tracing and DUA ontology and validation of our framework. Finally, we discuss our contributions and future works in the conclusion.



\section{Related Work}
In the previous research, we developed a Multi-aspect and Adaptive Trust-based Situation-aware access control framework for federated data-as-a-service systems (MATS) \cite{kim2022mats} and trusted compliance enforcement framework for sharing big data \cite{kim2021trusted}. The foundations of the frameworks are ontologies that represent trust and Health Insurance Portability and Accountability Act (HIPAA) \cite{hipaa, kim2021semantically} and query rewriting method that reasons over the ontologies \cite{oni2020framework}. The rationale behind the studies was to capture two aspects of the trust in big data exchange, which are imperative trust acquired by policy compliance \cite{kim2021trusted} and dynamic trust assessed by the reputation and behavior of users and organizations \cite{kim2022mats}.  However, we mainly focused on the trustworthiness of users and organizations participating in data exchange during the earlier research.

Thus, in this study, we focus on the trust elements of data. We categorized the trust elements of data in two: veracity and provenance. The categorization aligns with the notion of the trustworthiness of users and organizations. In this case, veracity is a dynamic element, and provenance is an imperative element. In this section, we illustrates achievements so far on veracity and their limitations. We will expand more on them in the following sections.

\subsection{Veracity}
Rubin and Lukoianova \cite{rubin2013veracity} surveyed the concept of big data veracity, which refers to the quality of big data and its potential biases, inaccuracies, and ambiguities. They proposed theoretical and empirical definitions of veracity and explored it across three main dimensions: objectivity, truthfulness, and credibility. The authors combined measures of these dimensions into a composite index to assess systematic variations in big data quality across datasets. They also categorized existing tools to measure veracity dimensions for the heterogeneous big data and identify critical information quality dimensions for each type.

However, Lozano et al. claims that most studies have a narrow scope and only focus on one type of data, indicating the need for more research to address the complex and multi-faceted problem of veracity assessment. They reviewed the current state of art regarding automatic veracity assessments and their methods used in social media and open-source data. The most common method is text analysis using supervised learning, with some use of deep learning.

Both forementioned papers could not clearly address evaluation criteria or the factors affects each categories of veracity. Rubin and Lukianova only proposed a value range of the composite index between 0 and 1 without any assessment method for three dimensions of objectivity, truthfulness, and credibility. Lozano et al. revealed that most research that they surveyed did not describe a complete assessment process.


On the contrary, Nurse et al. \cite{nurse2011information} provided a hierarchical framework for data quality that comprehensively reviews quality dimensions. The framework consists of four quality groups that address different aspects of data quality. The first group, \emph{intrinsic quality}, emphasizes that information has an inherent level of quality and includes dimensions such as accuracy, believability, objectivity, and reputation. The \emph{contextual quality} group considers the importance of assessing the context when judging quality and includes dimensions such as value-added, relevancy, timeliness, completeness, and appropriate amount of data. The \emph{representational quality} group focuses on data representation and includes dimensions such as interpretability, ease of understanding, representational consistency, and concise representation. The \emph{accessibility quality} group considers the ease with which desired information can be obtained and restricted as necessary, and includes dimensions such as accessibility and access security. Overall, this framework provides a comprehensive approach to assessing data quality in different aspects, and the dimensions defined can be used to guide quality assessment in practical applications. They also demonstrated the application of these factors in a real-life scenario and highlighted core influential factors and properties. They labeled types along with the factors - \emph{provenance, quality, and trustworthiness} - to categorize factors to look into for each assessment type. However, they did not provide ontology to show relationships between them.

There is no standard definition of veracity in data exchange. The diverse definitions include ''uncertainty due to data inconsistency, incompleteness, ambiguities, and deception," "truthfulness, accuracy or precision, correctness \cite{lozano2020veracity}," "objectivity, truthfulness, credibility \cite{rubin2013veracity}," and "the property that an assertion truthfully reflects the aspect it makes a statement about \cite{gollmann2012veracity}." Therefore, we use veracity as an umbrella term that includes trustworthiness, credibility and objectivity while following the definitions of each dimension claimed by Rubin and Lukoianova \cite{rubin2013veracity}.  One of the goal of this paper is to explore meaningful application of these concepts, though they does not have standardized meaning in data exchange. We will explore more on these concepts in subsequent subsections.

The Stanford Encyclopedia of Philosophy \cite{scientific} defines \emph{objectivity} as a value because it has certain importance or degrees that we approve of it. We expand the scope of the definition and assume that \emph{veracity} is a value because when we claim something objective, truthful, or credible, they implies certain importance or degrees that need consent or approval of others who share the notion of the subject. 

In this paper, we follow three main theoretical veracity dimentions proposed by Lukoianova \& Rubin \cite{rubin2013veracity}: objectivity, truthfulness, and credibility.

\subsubsection{Objectivity}
We assume \emph{objectivity} in 'subjective/objective' context. McQuail \cite{mcquail2010mcquail} claimed that \emph{objectivity} is a specific approach to the collection, processing, and dissemination of information that involves a particular attitude towards the task. It is concerned with both values and facts, and the facts themselves can have evaluative implications. Also, Haack \cite{haack2011defending} claimed that one way to differentiate science from the arts and other human endeavors is by its emphasis on objectivity, which is based on empirical evidence. This sets scientific knowledge apart from beliefs that are socially constructed and not necessarily grounded in factual reality.

According to McQuail, data objectivity is a specific approach to information processing and dissemination that involves a particular attitude towards the task, where both values and facts are taken into consideration, and the facts themselves can have evaluative implications. Haack further emphasizes the importance of empirical evidence in achieving data objectivity, stating that it distinguishes scientific knowledge from beliefs that are socially constructed and not necessarily grounded in factual reality. Thus, data objectivity requires a rigorous and systematic approach to the collection, processing, and dissemination of information, grounded in empirical evidence, and aimed at minimizing subjective biases and interpretations, by extending their points of view.

\subsubsection{Truthfulness}
In this paper, we focus the definition of "truthfulness" within the context of lying or deception, while recognizing its close association with other concepts such as faith, integrity, fact, and truth. As stated by Primoratz in his work on lying \cite{primoratz1984lying}, lying can be understood as the act of intentionally making a false statement with the aim of convincing someone else to believe it as true. Similarly, Isenberg \cite{isenberg1964deontology} defines lying as the creation of a false statement by an individual who does not believe it, with the intention of persuading another person to believe it as true. 

Nevertheless, we do not assume that data labeled with low truthfulness is necessarily the result of malicious intent. Thomas Carson's research \cite{carson2006definition, carson2010lying} on lying highlights that a speaker can present false or untruthful information to another person, without any intention of misleading them, under certain circumstances. In other words, this is the case when the speaker makes the statement within a context where they are expected to guarantee the truthfulness of their claims, and they believe themselves to be meeting this standard. Alternatively, the speaker may actually intend to warrant the truth of their statement, which can also lead to the presentation of false information.

\subsubsection{Credibility}

We follow Fogg and Tseng's perspective about credibility. They proposed trustworthiness and expertise as two critical components of credibility \cite{fogg1999elements}. We focused on the trustworthiness aspect among them since measuring expertise is out of our research scope. We limited the context of credibility to when the subject provides knowledge to others.

Trustworthiness as a credibility component differs from trustfulness illustrated in the previous section. Fogg and Tseng claim that trustworthiness is the perceived goodness or morality of the source \cite{fogg1999elements}. We translated their perspective as "If the data source provides data stated in the agreement."

\begin{figure*}[ht]
    \centering
    \includegraphics[width=0.6\textwidth]{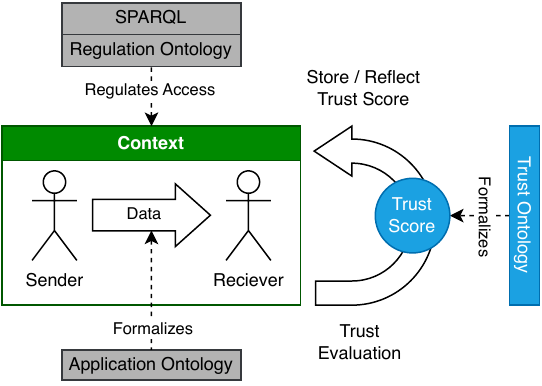}
    \caption{Trusted compliance enforcement framework. The framework grants access to data by reasoning over regulation ontology. According to users' compliance to the regulation, the framework evaluates their trust scores and updates them in the graph database.}
    \label{fig:tcef}
\end{figure*}
\begin{figure*}[!htb]
    \centering
    \includegraphics[width=\textwidth]{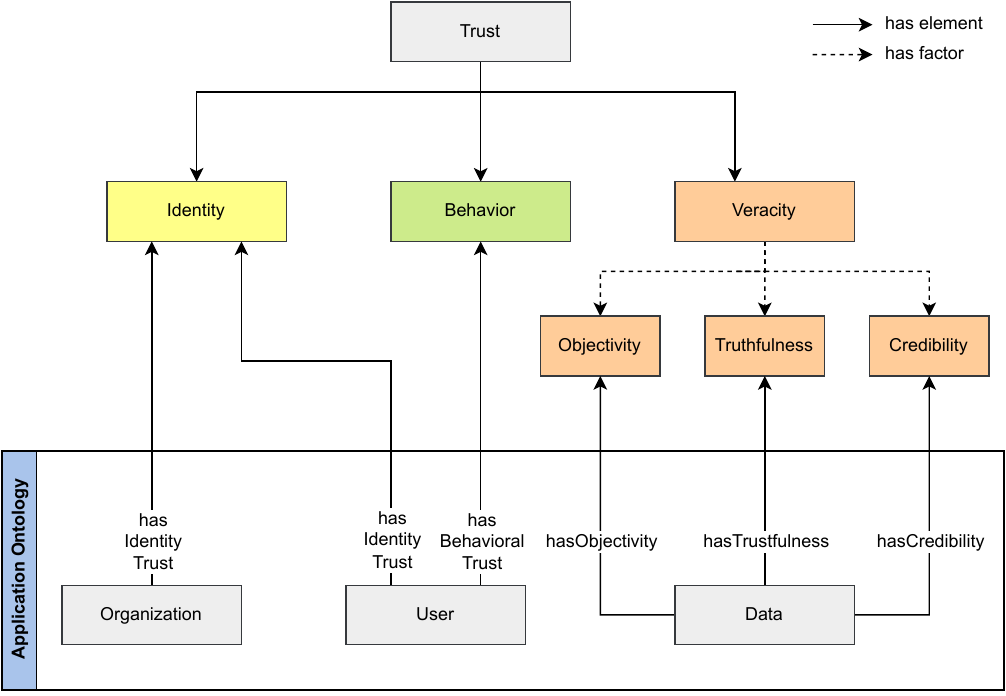}
    \caption{Trust ontology with three main pillars - identity, behavior, and veracity - and corresponding classes from an application ontology.}
    \label{fig:trust-knowledge-graph}
\end{figure*}

\section{Methods}
\subsection{TRUCE Framework}

We developed a TRUsted Compliance Enforcement service for secure health data exchange (TRUCE) to achieve the goal of this paper. We will explain the framework in a top-down manner to help understand the forest first and trees for more details. Figure \ref{fig:tcef} illustrates the components of the framework and ontologies required prior to real-world application at the most abstract level. In this subsection, we will briefly introduce each component and elaborate on them more in the following subsections.

\emph{Trust ontology} in the blue box formalizes the trust and consists of identity trust, behavioral trust, veracity, provenance, and their subclasses. The framework can identify the credentials of data exchange parties, information of the data, and the context of data exchange by reasoning over the \emph{data exchange ontology} and the \emph{trust ontology}. These ontologies are predefined part of the framework, while \emph{application ontology} and \emph{regulation ontology} is subject to change depands on use cases.

\emph{Application ontology} at the bottom of figure \ref{fig:tcef} formalizes the data. It defines the shape of the data and the relationships between its elements. \emph{Application ontology} is subject to change because every domain has different data types and shapes. We adopted COVID-19 Contact Tracing ontology for the use case of this paper.

\emph{Regulation ontology} is the ground truth of the trust evaluation mentioned earlier. The \emph{regulation ontology} defines relationships between stakeholders and articles in regulations that apply to a specific domain. SPARQL policy elaborates on the articles in the set of SPARQL queries. The framework grants or denies data exchange by reasoning over object and data properties of data, user, and organizations based on SPARQL policies and \emph{regulation ontology}. 

\begin{figure*}[ht]
    \centering
    \includegraphics[width=\textwidth]{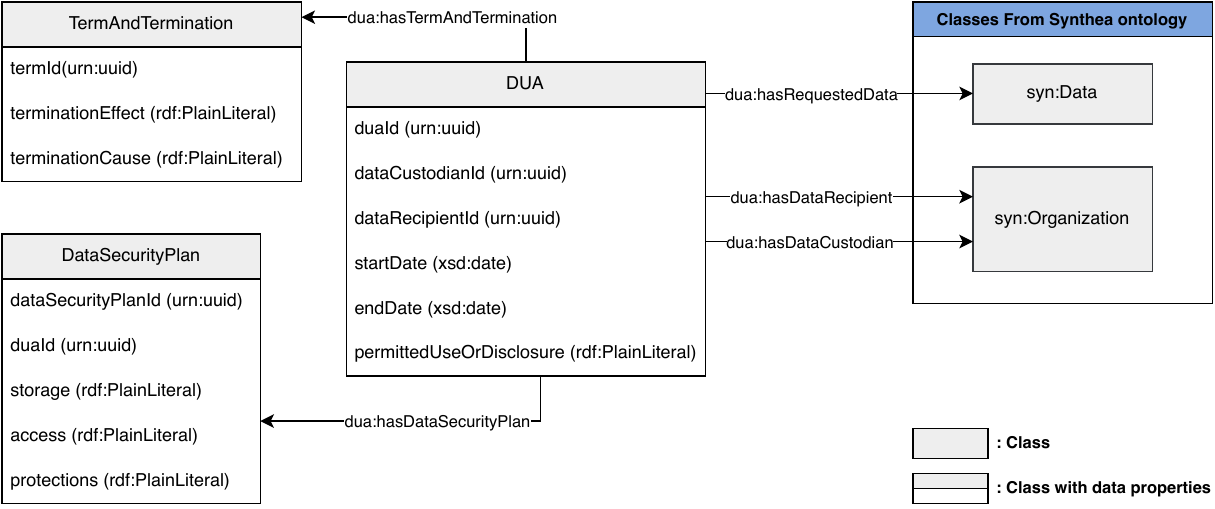}
    \caption{Data Usage Agreement Ontology}
    \label{fig:dua-ontology}
\end{figure*}

\subsection{Ontologies}

This section describes necessary ontologies of TRUCE framework: \emph{Trust}, \emph{CDC Contact Tracing}, and \emph{DUA}. In order to reconcile any overlapping classes between the ontologies, priority was given to the core classes of each ontology. Specifically, the classes \emph{Data} and \emph{Organization} from the \emph{CDC Contact Tracing} ontology replaced the corresponding classes in the \emph{DUA} ontology. Furthermore, the \emph{CDC Contact Tracing} ontology also superseded the \emph{Organization} and \emph{Data} classes from the \emph{Trust} ontology, while the data properties (\emph{Identity}, \emph{Behavior}, and \emph{Veracity}) of the \emph{Trust} ontology were modified to have the \emph{Data} and \emph{Organization} classes from the \emph{CDC Contact Tracing} ontology as their range. The ontologies are stored in the same graph database endpoint with different namespaces.


\begin{figure*}[ht]
    \centering
    \includegraphics[width=0.9\textwidth]{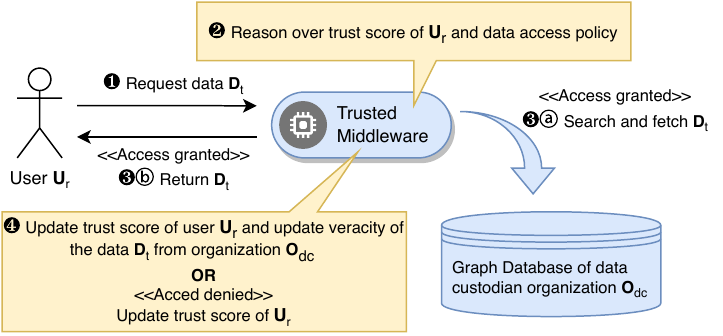}
    \caption{Trusted Middleware Data Exchange Cycle}
    \label{fig:tm-flowchart}
\end{figure*}

\subsubsection{Trust Ontology}

Figure \ref{fig:trust-knowledge-graph} illustrates trust ontology that we developed. The ontology is an extension of the trust ontology from our previous research \cite{kim2021trusted}. We categorized four elements of trust: \textit{behavioral trust, identity trust, data veracity}, which are in the green boxes. The hypothesis on each trust element is that we can observe each element and have their evaluation scores in advance before data exchange.

\textbf{Behavioral trust} denotes the estimated trustworthiness of users based on their actions on the data they receive. For example, if a user frequently breaches the regulations related to the data, we can say that the user has a low behavioral trust score. \textbf{Identity trust} represents the approximate trustworthiness of users based on their status. Illustrative factors affecting identity trust include a user's affiliation, role, and clearance level.

\textbf{Veracity} denotes the trustworthiness of the data. It has three factors, which are objectivity, truthfulness, and credibility. Veracity was part of our previous trust ontology but excluded from the application to COVID-19 contact tracing and Search and Rescue (SAR) use cases.

\subsubsection{Data Usage Agreement Ontology}

HIPAA mandates a covered entity (e.g. a healthcare provider, health plan, or healthcare clearing house) to enter into a written agreement with its business associates, which is known as a Business Associate Agreement (BAA). Data Usage Agreement (DUA) is the most common BAA document format. The document consists of following sections: \emph{permitted use or disclosure}, \emph{data security plan} (for storage, access and protections), \emph{term} and \emph{termination}. Therefore, organizations can implement appropriate administrative, physical, and technical safeguards to protect the confidentiality, integrity, and availability of the Protected Health Information (PHI).

The DUA ontology is depicted in Figure \ref{fig:dua-ontology}. It features the \texttt{DUA} class to encapsulate a DUA document and three classes that correspond to its constituent sections: \texttt{TermAndTermination}, \texttt{PermittedUseOrDisclosure}, and \texttt{DataSecurityPlan}. The \texttt{TermAndTermination} class establishes a relationship with the \texttt{DUA} class via the \texttt{hasPermittedUseOrDisclosure} relationship and comprises the data properties \texttt{term}, \texttt{terminationEffect}, and \texttt{terminationCause} with a range of \texttt{rdf:PlainLiteral}. Similarly, the \texttt{DataSecurityPlan} class connects to the \texttt{DUA} class via the \texttt{hasDataSecurityPlan} relationship and holds the data properties \texttt{storage}, \texttt{access}, and \texttt{protections} with a range of \texttt{rdf:PlainLiteral}. The \texttt{PermittedUseOrDisclosure} class, on the other hand, links to the \texttt{DUA} class through the \texttt{hasPermittedUseOrDisclosure} relationship and does not have any data properties. Instead, individuals of this class are utilized to represent permitted use or disclosure categories, as exemplified by the three diamond shapes in Figure \ref{fig:dua-ontology} (IRB approved research, public health, and health care operation).

The \texttt{Data} and \texttt{Organization} classes are sourced from the application ontology and respectively denote the data category requested by the recipient and the data custodian/recipient organization. These classes were incorporated from the application ontology to enhance the coherence and reusability of the DUA ontology. This separation of DUA-specific parts from the application ontology enables its application to different use cases as long as the application ontology contains the \texttt{Data} and \texttt{Organization} classes. In this manner, it is possible to reason over the eligibility of data exchange for specific categories.

\subsection{Trusted Middleware to Compute Trust and Veracity Scores}

Trusted Middleware (TM) manages data access, trust, and veracity scores on behalf of users \cite{kim2022mats,kim2021semantically,oni2020framework}. Users can access the data by TMs only, so we assume that TMs are tampering-proof and exclude other potential security issues.

Figure \ref{fig:tm-flowchart} illustrates the TM Data Exchange Cycle. As illustrated in the figure, TM reasons over trust scores and data access policy on behalf of users when they access specific data. For example, let us say that user $U_a$ requests data $D_t$. First, $U_a$ enters into the details of $D_t$ using the graphical interface of a TM. Then, the TM reasons over the trust score of $U_a$ and access policies related to $D_t$. The information that the access policies consider includes but is not limited to trust score, user organization, role, purpose, and classification level of the data.

When $U_a$ is qualified to access $D_t$, the TM grants access to $U_a$ and searches $D_t$ in its graph database and other organizations. Next, the TM aggregates $D_t$ from the organizations and returns the result to $U_a$. Then, the TM updates the trust score of $U_a$ and each organization's veracity score of $D_t$. Since $U_a$ followed policy and TM granted data access, the trust score of $U_a$ will remain unchanged or increase based on the trust evaluation method. Conversely, $U_a$ will assess each organization's veracity scores of $D_t$. Finally, TM propagates revaluated trust and veracity scores to other TMs in different organizations. 

On the other hand, when $U_a$ is unqualified, TM denies data access. Therefore, the trust score of $U_a$ will decrease or be the same depending on the tolerance parameter of the trust evaluation method.

TMs will reason over the updated user trust score and data veracity score during the following data exchange. Therefore, TMs will grant data exchange to trustworthy users and organizations on credible data based on accumulative trust and veracity evaluation. To summarize, TMs take charge of these processes on behalf of users and organizations: 

\begin{enumerate}
    \item Enforce data access regulation by reasoning over 
        \begin{itemize}
            \item Application ontology
            \item Trust ontology
            \item Data access policy
        \end{itemize}
    \item Fetching target data
    \item Update user trust score
    \item Update data veracity score
\end{enumerate}










\subsection{Trust Evaluation}
\subsubsection{Trust Evaluation Method}

The TRUCE framework evaluates trust scores based on a user's compliance with the data access policy. The trust policy refers to \emph{DUA} ontology and \emph{CDC Contact Tracing} ontology to grasp the context of rightful data access.

\begin{algorithm}
\caption{A trust initialization}\label{alg:trust-init}
\begin{algorithmic}
\Require $\mathbb{U}$ = users, \ $\mathbb{O}$ = oganizations, $\mathbb{T}$ = trust score
\For{$u_i \in \mathbb{U}$}
    \State $T_{Behavior}(u_i) \gets 1$
    \State $T_{Identity}(u_i) \gets 1$
\EndFor
\For{$o_i \in \mathbb{O}$}
    \State $T_{Credibility}(o_i) \gets 1$
    \State $T_{Identity}(o_i) \gets 1$
\EndFor
\end{algorithmic}
\end{algorithm}

Behavior trust score of users and identity trust score of users and organizations start from 1 in the [0, 1] score range, as illustrated in algorithm \ref{alg:trust-init}. The rationale behind the initial value is to support the paradigm shift from "need to share" to "need to know" while protecting data by setting a minimum trust score threshold.

\begin{algorithm}
\caption{An access control}\label{alg:trust-access}
\begin{algorithmic}
\Require $\mathbb{U}$  = user, $\mathbb{O}$  = organization, $\mathbb{D}$  = target data
\Statex
\Call {check compliance}{u,o,d}
\Statex
\Call {assess trust score }{u,o}
\If{compliant}
    \Statex \quad
    \Call {grant access}{}
\Else
    \Statex \quad
    \Call {deny access}{}
    \Statex \quad
    \Call {update trust score}{u, o}
\EndIf
\If{$O_{data\ custodian}(d)$ == $\varnothing$ }
    \Statex \quad
    \Call {notice user}{u}
    \Statex \quad
    \Call {update credibility}{$O_{data\ custodian}$}
\EndIf
\If{$T_{Credibility}(o) \  \leq \ 0$ or $T_{Identity}(o) \  \leq \ 0$}
    \Statex \quad
    \Call {rewrite data usage agreement}{}
\EndIf
\end{algorithmic}
\end{algorithm}

In our validation settings, all trusted middleware has no errors and is reliable. Algorithm \ref{alg:trust-access} describes data access control when a user requests data from a trusted middleware. First, it checks the user's compliance with the data access policy, which is DUA. Then it assesses the trust score of the user. Based on the results of the two processes, trusted middleware grant or deny data access. If the user is not compliant, trust middleware will lower the user's and organization's trust scores. When the user is compliant, trusted middleware checks if the data custodian organization has target data. If the organization does not have target data, trusted middleware lowers the credibility score of the organization because the user appropriately requested data under the DUA, but the organization does not have data specified in the DUA.

\begin{algorithm}
\caption{A trust assessment}\label{alg:trust-assessment}
\begin{algorithmic}
\Require $\mathbb{W}$ = weight, $\mathbb{T}$ = trust score, $\mathbb{U}$ = user
\State $T_{weighted\ average} \gets$ \par $W_{Behavior} \cdot T_{Behavior}(u)\ + \ W_{Identity} \cdot T_{Identity}(u)$
\If{$T_{weighted\ average} \geq T_{threshold}$}
    \Statex \quad
    \Call {return}{True}
\Else
    \Statex \quad
    \Call {retur}{False}
\EndIf
\end{algorithmic}
\end{algorithm}

Algorithm \ref{alg:trust-assessment} describes trust assessment. Trusted middleware calculates a weighted average of a user's behavior and identity trust. A data custodian organization sets weights for each trust score and a threshold. Trusted middleware grants data access if the weighted average is equal to or greater than the threshold.

When a data custodian organization's credibility score becomes zero or the data requester organization's identity trust score becomes zero, trusted middleware prevents further data exchanges between the organizations until they rewrite DUA, as illustrated in algorithm \ref{alg:trust-access}.

\subsubsection{Policy Examples}

\begin{lstlisting}[
  language=sparql,
  caption={Data access policy to check the existence of DUA between the user and the data custodian organization.},
  label={policy:dua-existence},
  captionpos=b
]
ASK{
   ?dataCustodian a syn:Organization . 
   ?dataCustodian rdfs:label "DataCustodian"^^rdf:PlainLiteral . 
   ?user a tst:User . 
   ?user rdfs:label "physician_105"^^rdf:PlainLiteral . 
   ?user syn:isAffiliatedWith ?organization . 
   ?dua a dua:DataUsageAgreement . 
   ?dua dua:hasRecipient ?organization . 
   ?dua dua:hasDataCustodian ?dataCustodian . 
}
\end{lstlisting}

SPARQL query \ref{policy:dua-existence} is a data access policy to check if a user's affiliated organization has made DUA with the data custodian organization. First, the policy identifies the \emph{physician\_105}'s organization in this example. Then, it looks for a \emph{dua:DataUsageAgreement} instance, which has the \emph{?organization} as a data recipient and the \emph{?dataCustodian} as a data custodian. Since the policy is a SPARQL ASK query, it will return a boolean value \emph{true} when there is a DUA between the organizations. Otherwise, it will return \emph{false}, which means that the user violated the data access policy - attempting access to data without a data usage agreement.

\begin{lstlisting}[
  language=sparql,
  float,
  caption={Data access policy to check the existence of the requested data in the DUA.},
  label={policy:requested-data},
  captionpos=b
]
ASK{
   ?dataCustodian a syn:Organization . 
   ?dataCustodian rdfs:label "DataCustodian"^^rdf:PlainLiteral . 
   ?user a tst:User . 
   ?user rdfs:label "nurse_207"^^rdf:PlainLiteral . 
   ?user syn:isAffiliatedWith ?organization . 
   ?dua a dua:DataUsageAgreement . 
   ?dua dua:hasRecipient ?organization . 
   ?dua dua:hasDataCustodian ?dataCustodian . 
   ?dua dua:requestedData syn:Patient^^rdf:PlainLiteral . 
}
\end{lstlisting}

SPARQL query \ref{policy:requested-data} is a data access policy that looks into the detail of the DUA between the organizations. This policy assumes that the \emph{nurse\_208} passed the policy \ref{policy:dua-existence} and sees if the user requested one of the data categories stated in the DUA. In this case, the nurse requested a patient data category. Similar to the policy \ref{policy:dua-existence}, it identifies the \emph{?dua} between the organizations and see if \emph{?dua} has \emph{syn:Patient} as \emph{dua:requestedData}. When it is \emph{true}, TM grants access to the patient data or denies access otherwise.

\begin{lstlisting}[
  language=sparql,
  float,
  caption={Data custodian policy to check the existence of the data category requested in the DUA between the organizations.},
  label={policy:dua-data-category},
  captionpos=b
]
ASK {
  ?dataCustodian a syn:Organization .
  ?dataCustodian rdfs:label "DataCustodian"^^rdf:PlainLiteral .
  ?user a tst:User .
  ?user rdfs:label "nurse_629"^^rdf:PlainLiteral .
  ?user syn:isAffiliatedWith ?org .
  ?dua a dua:DataUsageAgreement .
  ?dua dua:hasRecipient ?org .
  ?dua dua:hasDataCustodian ?dataCustodian .
  ?dua dua:requestedData ?requestedData.
  FILTER(STR(?requestedData) IN ( STR(syn:Encounter), STR(syn:Observation), STR(syn:Patient)))
}
\end{lstlisting}

\begin{table*}[ht]
\centering
\resizebox{0.7\textwidth}{!}{%
\begin{tabular}{|l|c|c|c|c|}
\hline
Execution time (seconds)        & 1K      & 10K     & 100K    & 1M       \\ \hline
Recipient policy check          & 0.01416 & 0.01398 & 0.01714 & 0.01708  \\ \hline
Data credibility check          & 0.00651 & 0.00680 & 0.00856 & 0.00699  \\ \hline
Trust score update              & 0.02299 & 0.05159 & 0.03807 & 0.03373  \\ \hline
Data retrieval                  & 0.03176 & 0.38014 & 2.30240 & 24.54431 \\ \hline
\end{tabular}%
}
\captionsetup{font=footnotesize}
\caption{Average data retrieval time from 1,000 transactions. Each transaction has different user attributes, e.g., their organization and details about its DUA with the data custodian organization.}
\label{tab:result}
\end{table*}

SPARQL query \ref{policy:dua-data-category} checks the data custodian has data category requested in a DUA between organizations. The first half of the query is similar to SPARQL query \ref{policy:requested-data}, but in this case, the data custodian is expected to have a requested data category (`dua:requestedData`) matching the value of `?requestedData`. The FILTER function is used to filter the results, checking if the `?requestedData` matches any of the specified data categories: `syn:Encounter`, `syn:Observation`, or `syn:Patient`. If the query returns true, it means that the data custodian, represented by `?dataCustodian`, possesses the requested data categories as specified in the DUA between organizations. If the query returns false, it indicates that the data custodian does not have the required data categories in the DUA.

\begin{lstlisting}[
  language=sparql,
  float,
  caption={Query to update the behavior trust of the user. The query reduces the behavior trust of user 105 assuming that the user violated DUA policy.},
  label={policy:behavior-update},
  captionpos=b
]
DELETE {
   ?user tst:behaviorTrust "1.0"^^xsd:float . 
}
INSERT {
   ?user tst:behaviorTrust "0.9"^^xsd:float . 
}
WHERE {
   ?user a tst:User . 
   ?user rdfs:label "research_scientist_731"^^rdf:PlainLiteral . 
}
\end{lstlisting}

SPARQL query \ref{policy:behavior-update} updates the behavior trust of a user. This case assumes that the \emph{research\_scientist\_731} violated a DUA policy and reduces the behavior trust score of the user. The data custodian organization can set a weight for each violation. In this example, we set 0.1 as a weight; therefore, user 105's behavior trust score decreased to 0.9.



\subsection{Dataset}

\begin{figure*}[ht]
    \centering
    \includegraphics[width=\textwidth]{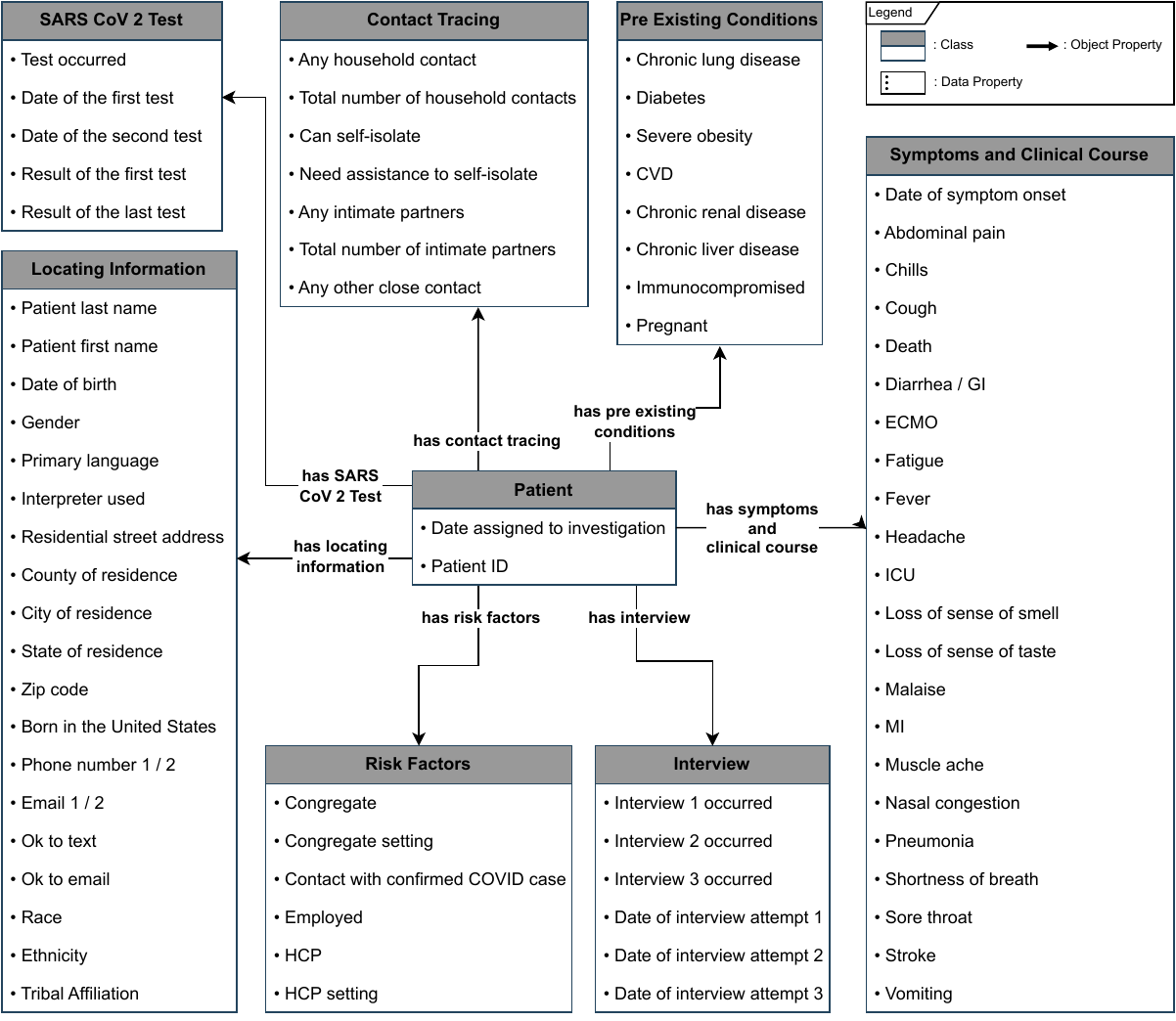}
    \caption{CDC Contact Tracing Ontology}
    \label{fig:cdc-ontology}
\end{figure*}

We have two data generation stages - data generation and data conversion. First, we generated a synthetic dataset of up to one million records of COVID-19 patients to validate the framework. We developed a comprehensive dataset utilizing the \emph{CDC Contact Tracing} ontology illustrated in figure \ref{fig:cdc-ontology} \cite{kim2021trusted, contacttracing}, with the Patient class as its core component. The dataset encompasses several classes, addressing various aspects of patient information in relation to SARS-CoV-2, including test results, contact tracing, pre-existing conditions, symptoms and clinical course, interview data, risk factors, and locating information.

Given that the CDC provided the essential contact tracing data in table format, we defined corresponding data and object properties to facilitate the storage of this information in a graph database, thereby adding contextual richness. Moreover, we incorporated organization data, detailing their DUA contracts and the assigned data custodian, and generated user data for each organization. Trust scores were assigned to both organizations and individual users, enhancing the reliability of the dataset.

This structured and enriched dataset, developed with a focus on technical robustness and credibility, aims to serve as a valuable resource for further research in this domain, aligning with the technical orientation of a computer science conference.




To simulate data exchange, we implemented TM server using the Python Flask web application framework \cite{grinberg2018flask}. The TM server features a Graphical User Interface (GUI) presented through a HTML document rendered using the Jinja2 template engine and endpoints for simulation automation. 

For storage and processing of RDF data, we utilized Apache Jena \cite{jena} and for the SPARQL Graph Store protocol, we employed Fuseki \cite{fuseki}. Typically, Apache Jena and Fuseki are packaged together as Apache Jena Fuseki (Fuseki). We utilized Docker Compose v2 to deploy the TM and Fuseki servers in separate Docker containers within the same network. To define the relationships between these components in the Docker Compose file, we built a Docker image for the TM and the Jena Fuseki Docker image \cite{leroy}  based on Stain Soiland-Reyes \cite{stain} for reproducibility. Since our graph database has huge size, we uploaded the dataset by Jena Fuseki API instead of its GUI interface.

\begin{figure}[ht]
    \centering
    \includegraphics[width=\columnwidth]{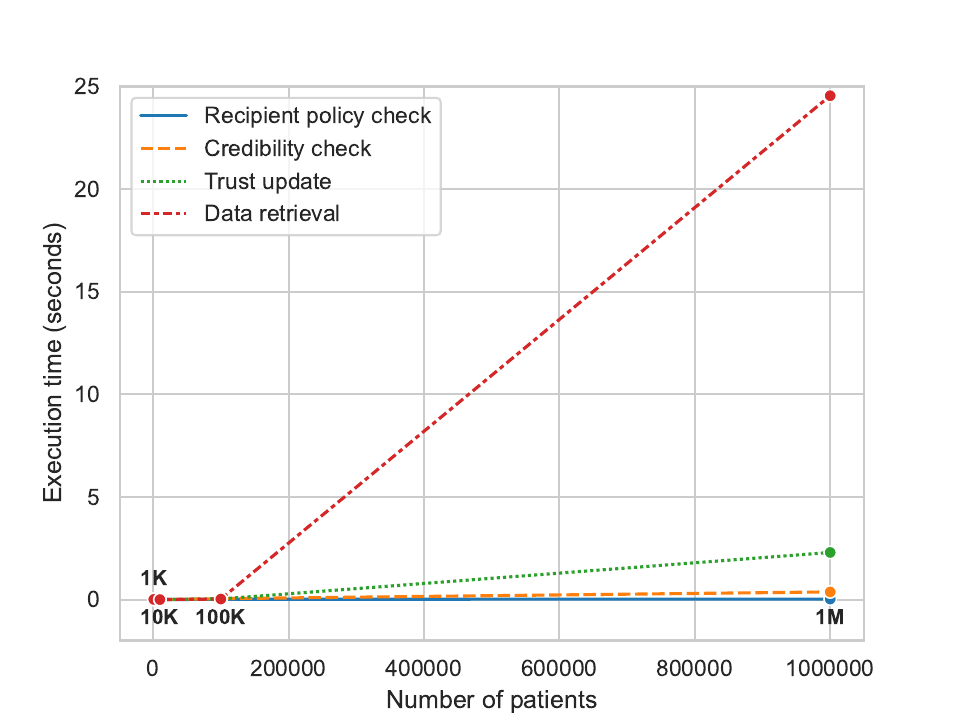}
    \caption{Execution time of each measurement on dataset sizes from 1k to 1m. Query execution time linearly increased according to the dataset size while policy check and trust update time have little change.}
    \label{fig:execution-time}
\end{figure}

\section{Result}

\subsection{Time Complexity}

We validated the TRUCE framework in four datasets - 1K, 10K, 100K, and 1M synthetic patients data. The purpose of the validation was to confirm the practicality of the framework by measuring the execution time of significant stages during the data exchange: recipient policy check, data credibility check, trust and credibility score update, and target data retrieval.

We generated a hundred users under ten organizations. Each organization has a different DUA with the same data custodian organization, and each DUA has different requested data and permitted usage or disclosure. The demographics of organizations are as follows: Seven organizations have written DUA with the data custodian organization. Among them, four organizations have requested \emph{patient} data class in their DUA. Half of the organizations who requested \emph{patient} data specified \emph{public health} as \emph{permitted use or disclosure}. In this setting, the users requested \emph{patient} data for \emph{public health}. The trusted middleware granted data access based on their organization's DUA information.

Table \ref{tab:result} describes the validation results. It turned out that the recipient policy check and data credibility check stages had no significant impact on the data exchange. The trust and credibility score update stage had more impact than the previous stages but had no crucial impact.

Figure \ref{fig:execution-time} illustrates the results more intuitively. The trust and credibility check executions have O(1) time complexity. The trust and credibility update has time complexity similar to O(logn). On the other hand, data retrieval has a time complexity similar to $O(n^2)$.

\subsection{Trust and Behavior Score Trajectory}

In this study, we conducted an experiment to examine the behavior trust score of users and the credibility score of organizations during data exchange in various contexts. Our objective was to investigate the dynamics of trust and credibility in different scenarios, shedding light on the impact of data usage agreements (DUA), and the alignment of data categories and properties.

\begin{figure}[ht]
    \centering
    \includegraphics[width=\columnwidth]{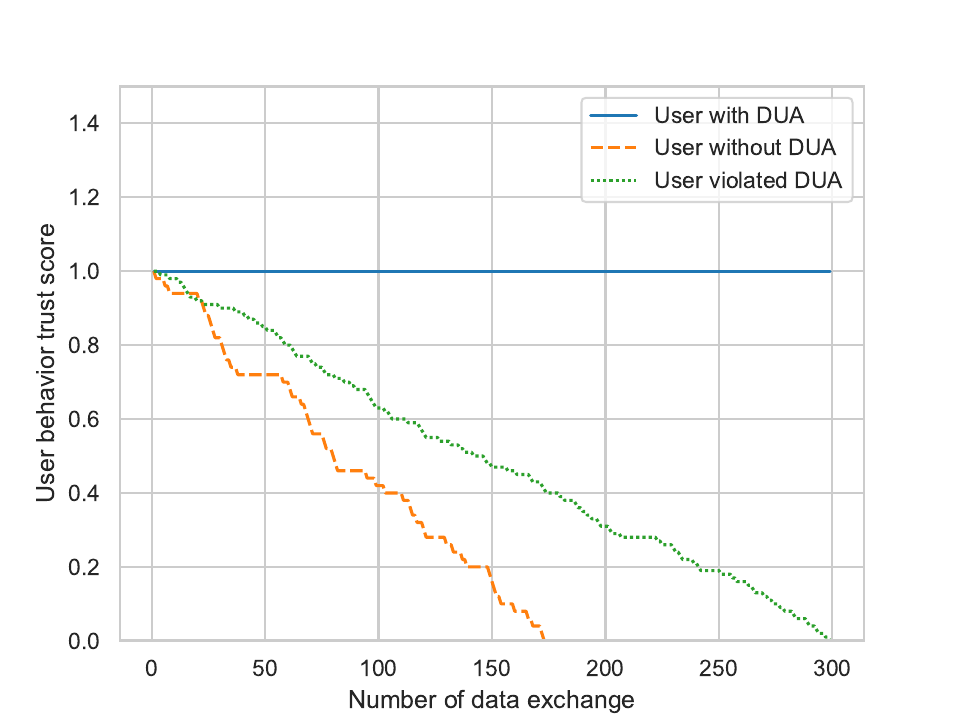}
    \caption{A behavior trust trajectory of users with three different DUA contexts. The users without DUA and who violated DUA requested illegitimate request with 30 percent of chance during the experiment.}
    \label{fig:usef-trust-score}
\end{figure}

\begin{figure}[ht]
    \centering
    \includegraphics[width=\columnwidth]{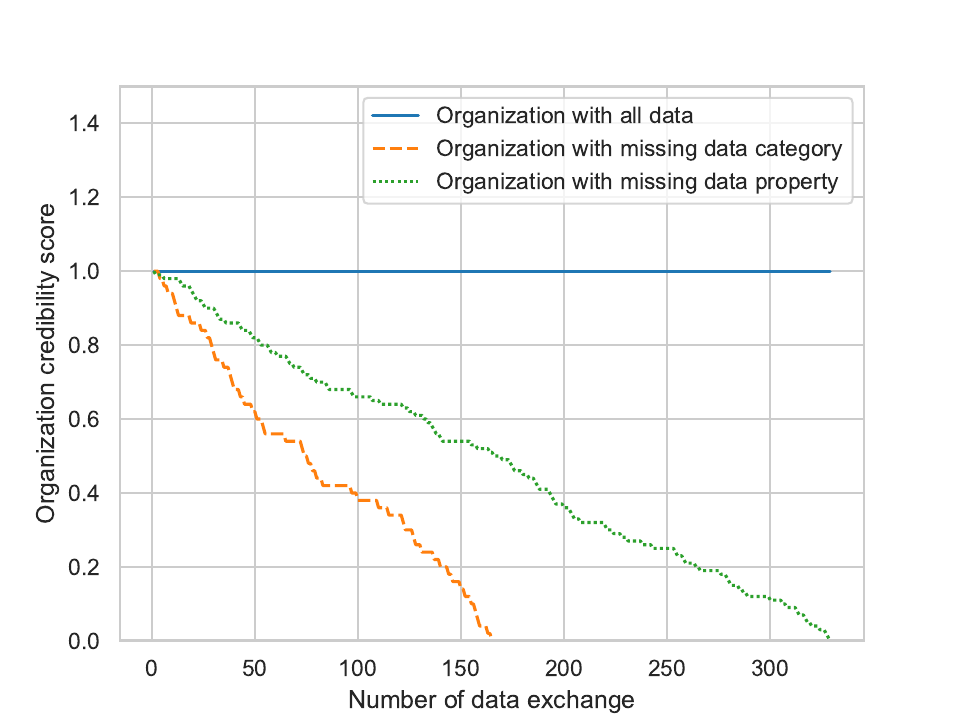}
    \caption{A credibility score trajectory of organizations with three different DUA contexts. The organizations without DUA-requested data category and without complete data had 30 percent of chance to get request regarding the data. }
    \label{fig:org-credibility-score}
\end{figure}

One of the key aspects of our experiment was the tracking of the behavior trust score of users. The initial behavior trust score and credibility for both users and organizations involved in the data exchange process were set to 1. We aimed to understand how users' trust in the data custodian organization varied across different contexts. Specifically, we considered three distinct scenarios for the behavior trust score of users. First, we analyzed users affiliated with organizations that had a DUA with the data custodian organization. In this scenario, violations of the DUA occurred with a 30\% chance for each user until their behavior trust score reached zero. In the event that a user violated the DUA, we deducted 0.01 from their behavior trust score. In case of the user affiliated with organizations that did not have a DUA with the data custodian organization. In this case, a higher deduction of 0.02 was implemented for users without a DUA, reflecting the potentially increased risk associated with data exchange without a formal agreement.

Additionally, we explored the credibility score of the data custodian organization in different contexts. We focused on three scenarios for the organization's credibility score, the same 30\% chance of violations was applied. First, we investigated situations where the data custodian organization had all the data categories specified in the DUA with the data requesting organization. If the data custodian did not possess the data category requested in the DUA, a deduction of 0.02 was made from their credibility score. Second, we examined cases where the data custodian organization had all the data categories specified in the DUA but had missing data properties within the requested data category. In this instance, a deduction of 0.01 was applied to their credibility score to account for the incomplete data properties.

By examining both the behavior trust scores of users and the credibility scores of organizations in these diverse contexts, we sought to uncover valuable insights into the dynamics of trust and credibility during data exchange. This research has significant implications for understanding the factors that contribute to successful data collaborations and establishing effective mechanisms for ensuring trust and credibility in such exchanges.

\section{Conclusion}
In this research, we developed a TRUsted Compliance Enforcement service for secure health data exchange (TRUCE) framework in this research. We applied the TRUCE framework to the healthcare domain with HIPAA DUA as a ground truth. Our experiment showed that it is possible to grasp the context of data exchange and dynamically manage the trust score of users and organizations. Also, we measured data retrieval time on different dataset sizes from 1K to 1M and found that regulatory compliance does not impose a considerable performance overhead.

We made contributions as follows. First, we expanded our previous research \cite{kim2021semantically,kim2021trusted} and developed a concrete trust management framework with firm ground truth. The framework reasons over details of HIPAA DUA between organizations and provides data access according to the context of the data exchange.

Second, we developed HIPAA DUA ontology to grasp the contexts of data exchange and protect health data accordingly. The ontology comprises five classes illustrating mandatory HIPAA requirements before data exchange between organizations. Therefore, it was possible to catch the context of the data exchange. For example, we can check if a user requests data that falls into the categories specified in a DUA. Also, it is possible to confirm the purpose of data exchange by reasoning over permitted use or disclosure specified in the DUA.

Third, we introduced veracity to our trust ontology \cite{kim2022mats}. As a result, the TRUCE framework can achieve bidirectional trust management since the direction of credibility is from the data recipient to the data custodian and the data.

Finally, we validated our framework and ontologies to real-world scenarios. The scenario assumes that one organization provides health data to users from other organizations with or without DUA, and the TRUCE framework works as an access control and trust management authority. To populate health data, we adopted CDC Contact Tracing ontology we develope from our previous research \cite{kim2021trusted, contacttracing}.

However, there are some limitations of our research. The target trust components of the validations were identity trust and credibility only because they are the only components that the TRUCE framework can judge by reasoning over DUA ontology. It was possible to check user violations by policies referring to DUA ontology. Also, it was possible to measure credibility by reasoning over \emph{dua:hasRequestedData} properties of the DUA between the organizations.

On the other hand, the remaining trust components - identity trust, objectivity, and truthfulness - require human intervention factors. For example, in the case of identity trust, there must be an offset identity trust against each organization from their knowledge about it. We should also consider reputations and recommendations from other organizations. Moreover, there can be a bias in each knowledge, reputation, and recommendation.

Secondly, applying our framework and ontologies to other use cases in different situations is necessary. That says, we keep open-world assumptions in mind and know there are still many yet-investigated contexts of trust management and its application to real-world domains.

Our future work will explore novel solutions to the limitations of our research mentioned above. Also, we will validate the framework with more data access policy to more precisely grasp the contexts of health data exchange. Furthermore, we will expand our previous research \cite{kim2022mats} on trust management in federated data exchange by game theory to health data exchange.

\section*{Acknowledgments}
This research was partially supported by NSF award 1747724, Phase I IUCRC UMBC: Center for Accelerated Real time Analytics (CARTA).

\bibliographystyle{IEEEtran}
\bibliography{reference}

@article{kim2022mats,
  title={MATS: A Multi-aspect and Adaptive Trust-based Situation-aware Access Control Framework for Federated Data-as-a-Service Systems},
  author={Kim, Dae-young and Alodadi, Nujood and Chen, Zhiyuan and Joshi, Karuna and Crainiceanu, Adina and Needham, Don},
  journal={UMBC Student Collection},
  year={2022}
}

@inproceedings{oni2020framework,
  title={A Framework for Situation-Aware Access Control in Federated Data-as-a-Service Systems Based on Query Rewriting},
  author={Oni, Samson and Chen, Zhiyuan and Crainiceanu, Adina and Joshi, Karuna P and Needham, Don},
  booktitle={2020 IEEE International Conference on Services Computing (SCC)},
  pages={1--11},
  year={2020},
  organization={IEEE}
}

@inproceedings{kim2021semantically,
  title={A semantically rich knowledge graph to automate hipaa regulations for cloud health it services},
  author={Kim, Dae-young and Joshi, Karuna P},
  booktitle={2021 7th IEEE Intl Conference on Big Data Security on Cloud (BigDataSecurity), IEEE Intl Conference on High Performance and Smart Computing,(HPSC) and IEEE Intl Conference on Intelligent Data and Security (IDS)},
  pages={7--12},
  year={2021},
  organization={IEEE}
}

@inproceedings{kim2021trusted,
  title={Trusted Compliance Enforcement Framework for Sharing Health Big Data},
  author={Kim, Dae-young and Elluri, Lavanya and Joshi, Karuna P},
  booktitle={2021 IEEE International Conference on Big Data (Big Data)},
  pages={4715--4724},
  year={2021},
  organization={IEEE}
}

@misc{hipaa,
  author = {{US Department of Health and Human Services (HHS) }},
  howpublished = {https://www.hhs.gov/hipaa/index.html},
  title = {{The Health Insurance Portability and Accountability Act (HIPAA)}},
  year = 1996
}

@article{lozano2020veracity,
  title={Veracity assessment of online data},
  author={Lozano, Marianela Garc{\'\i}a and Brynielsson, Joel and Franke, Ulrik and Rosell, Magnus and Tj{\"o}rnhammar, Edward and Varga, Stefan and Vlassov, Vladimir},
  journal={Decision Support Systems},
  volume={129},
  pages={113132},
  year={2020},
  publisher={Elsevier}
}

@inproceedings{hardalov2016search,
  title={In search of credible news},
  author={Hardalov, Momchil and Koychev, Ivan and Nakov, Preslav},
  booktitle={International conference on Artificial intelligence: methodology, systems, and applications},
  pages={172--180},
  year={2016},
  organization={Springer}
}

@article{igawa2016account,
  title={Account classification in online social networks with LBCA and wavelets},
  author={Igawa, Rodrigo Augusto and Barbon Jr, Sylvio and Paulo, K{\'a}tia Cristina Silva and Kido, Guilherme Sakaji and Guido, Rodrigo Capobianco and J{\'u}nior, Mario Lemes Proen{\c{c}}a and da Silva, Ivan Nunes},
  journal={Information Sciences},
  volume={332},
  pages={72--83},
  year={2016},
  publisher={Elsevier}
}

@article{rubin2013veracity,
  title={Veracity roadmap: Is big data objective, truthful and credible?},
  author={Rubin, Victoria and Lukoianova, Tatiana},
  journal={Advances in Classification Research Online},
  volume={24},
  number={1},
  pages={4},
  year={2013}
}

@inproceedings{gollmann2012veracity,
  title={Veracity, plausibility, and reputation},
  author={Gollmann, Dieter},
  booktitle={IFIP International Workshop on Information Security Theory and Practice},
  pages={20--28},
  year={2012},
  organization={Springer}
}

@book{grinberg2018flask,
  title={Flask web development: developing web applications with python},
  author={Grinberg, Miguel},
  year={2018},
  publisher={" O'Reilly Media, Inc."}
}

@misc{jena,
    author = {{Apache Software Foundation}},
    title = {Apache Jena},
    year = {2021},
    howpublished = {\url{https://jena.apache.org/}}
}

@misc{fuseki,
    author = {{Apache Software Foundation}},
    title = {Apache Jena Fuseki},
    howpublished = {\url{https://jena.apache.org/documentation/fuseki2}}
}

@inproceedings{ribeiro2016should,
  title={"Why should i trust you?" Explaining the predictions of any classifier},
  author={Ribeiro, Marco Tulio and Singh, Sameer and Guestrin, Carlos},
  booktitle={Proceedings of the 22nd ACM SIGKDD international conference on knowledge discovery and data mining},
  pages={1135--1144},
  year={2016}
}

@inproceedings{nurse2011information,
  title={Information quality and trustworthiness: A topical state-of-the-art review},
  author={Nurse, Jason RC and Rahman, Syed Sadiqur and Creese, Sadie and Goldsmith, Michael and Lamberts, Koen},
  year={2011},
  booktitle={The International Conference on Computer Applications and Network Security (ICCANS) 2011},
  publisher={IEEE}
}

@misc{stain,
    author = {Stian Soiland-Reyes},
    title = {stain/jena-fuseki},
    howpublished = {\url{https://hub.docker.com/r/stain/jena-fuseki}}
}

@misc{leroy,
    author = {Dae-young Kim},
    title = {leroykim/jena-fuseki-4.7.0},
    howpublished = {\url{https://hub.docker.com/repository/docker/leroykim/jena-fuseki-4.7.0}}
}

@misc{scientific,
    author = {{Stanford Encyclopedia of Philosophy}},
    title = {Scientific Objectivity},
    howpublished = {\url{https://plato.stanford.edu/entries/scientific-objectivity/}}
}

@book{mcquail2010mcquail,
  title={McQuail's mass communication theory},
  author={McQuail, Denis},
  year={2010},
  publisher={Sage publications}
}

@book{haack2011defending,
  title={Defending science-within reason: Between scientism and cynicism},
  author={Haack, Susan},
  year={2011},
  publisher={Prometheus Books}
}

@article{primoratz1984lying,
  title={Lying and the “Methods of Ethics”},
  author={Primoratz, Igor},
  journal={International Studies in Philosophy},
  volume={16},
  number={3},
  pages={35--57},
  year={1984}
}

@article{isenberg1964deontology,
  title={Deontology and the Ethics of Lying},
  author={Isenberg, Arnold},
  journal={Philosophy and Phenomenological Research},
  volume={24},
  number={4},
  pages={463--480},
  year={1964},
  publisher={JSTOR}
}

@article{carson2006definition,
  title={The definition of lying},
  author={Carson, Thomas L},
  journal={No{\^u}s},
  volume={40},
  number={2},
  pages={284--306},
  year={2006},
  publisher={Wiley Online Library}
}

@book{carson2010lying,
  title={Lying and deception: Theory and practice},
  author={Carson, Thomas L},
  year={2010},
  publisher={OUP Oxford}
}

@inproceedings{fogg1999elements,
  title={The elements of computer credibility},
  author={Fogg, Brian J and Tseng, Hsiang},
  booktitle={Proceedings of the SIGCHI conference on Human Factors in Computing Systems},
  pages={80--87},
  year={1999}
}

@MISC{cdcguide,
author = {{Centers for Disease Control and Prevention (CDC)}},
title = {Interim Guidance on Developing a COVID-19 Case Investigation \& Contact Tracing Plan},
month = Feb,
year = {2022},
howpublished={\url{https://archive.cdc.gov/\#/details?url=https://www.cdc.gov/coronavirus/2019-ncov/php/contact-tracing/contact-tracing-plan/overview.html}},
note = {Accessed: 2023-09-05}
}

@misc{contacttracing,
    author = {{Dae-young} Kim},
    title = {CDC Contact Tracing Ontology},
    year = {2023},
    publisher = {GitHub},
    journal = {GitHub repository},
    howpublished = {\url{https://github.com/leroykim/cdc-contact-tracing-ontology}}
}

\begin{IEEEbiography}[{\includegraphics[width=1in,height=1.25in,clip,keepaspectratio]{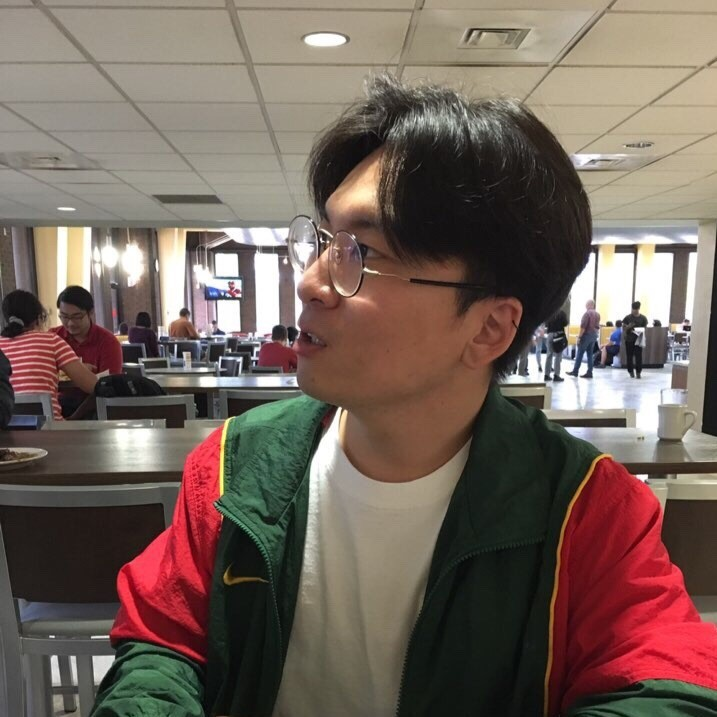}}]{{Dr. Dae-young Kim}}
holds a Ph.D. in Information Systems and was affiliated with the Knowledge Analytics Cognitive and Cloud (KnACC) Lab. His cutting-edge research is centered on the application of the Semantic Web in health informatics. Specifically, Dr. Kim delves deep into healthcare regulation automation and the intricate facets of trust management between data exchange parties, ensuring alignment with the `need to know' paradigm. In addition to this academic endeavor, he possesses valuable industry experience, notably in representing smart manufacturing data using semantic web technology.
\end{IEEEbiography}

\vskip 0pt plus -1fil

\begin{IEEEbiography}[{\includegraphics[width=1in,height=1.25in,clip,keepaspectratio]{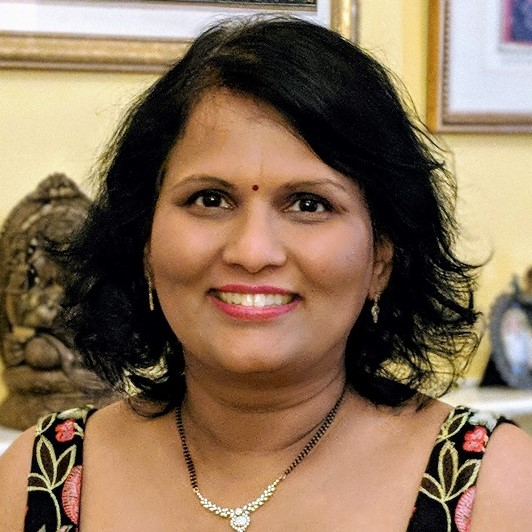}}]{{Dr. Karuna Pande Joshi}}
is a Professor of
Information Systems at the University of Maryland, Baltimore County (UMBC) and UMBC 
Director of Center for Accelerated Real Time Analytics (CARTA). She also directs the Knowledge
Analytics Cognitive and Cloud (KnACC) Lab. Her
research focus is in the areas of Data Science,
Cloud Computing, Data Compliance, and Healthcare IT Systems. She has published over 80 papers, and her research is supported by NSF, ONR, DoD, GE Research, and Cisco. She teaches courses in Big Data, Database Design, and Software Engineering. She received her MS and Ph.D. in Computer Science from UMBC, where she was twice awarded the IBM Ph.D. Fellowship, and her Bachelors in Computer Engineering from the University of Mumbai, India. Dr. Joshi also has extensive experience working in the industry, primarily as an IT Program/Project Manager at the International Monetary Fund.
\end{IEEEbiography}

\end{document}